\documentstyle[preprint,prb,aps]{revtex}
\begin{document}
\draft
\title{Exact shape of the lowest Landau level in a double--layer system
and a superlattice with uncorrelated disorder}

\author{T.~V.~Shahbazyan$^{(1)}$ and M.~E.~Raikh$^{(2)}$}
\address{$^{(1)}$ Department of Physics and Astronomy, 
Ohio University, Athens, Ohio 45701\\
$^{(2)}$ Department of Physics, University of Utah, 
Salt Lake City, Utah  84112}

\maketitle
\begin{abstract}
We extend Wegner's exact solution for the 2D 
density of states at  the  lowest Landau level with a short--range
disorder to the cases of a double--layer system and a superlattice.
For the double--layer system, an analytical expression for the
density of states, illustrating the interplay between the tunnel
splitting of Landau levels and the disorder--induced broadening, is
obtained. For the superlattice, we derive an integral equation, the
eigenvalue of which determines the exact density of states. By
solving this equation numerically, we trace the disappearance of the
miniband with increasing disorder.     

\end{abstract}
\pacs{PACS numbers: 73.20.Dx, 73.40.Hm}

\narrowtext

The shape of the Landau levels (LL) in a 2D system in the presence of
a disorder was the subject of intensive study during the last two
decades.\cite{A,B,C,D,E,F,G,H,I,J,K,L,M,N,O,P,Q,R,S}
The complexity of the problem arises from the fact that in the absence
of the disorder the energy spectrum is discrete.  As a result, 
the self--energy of an electron appears to be real in any finite order
of the perturbation theory. Therefore, obtaining a finite width of the
LL requires summation of the entire diagram expansion. It was
demonstrated\cite{J,P} that such a summation is possible when the
number of the  LL is large. The simplifications, arising in this
limit, are different in the case of a short--range and a  smooth
disorder. In the former case only a subsequence  of diagrams without
self--intersections contributes to the self--energy, or, in other
words, the self--consistent Born approximation \cite{A,C} becomes 
asymptotically exact.\cite{K} The shape of the LL in this case is
close to semielliptical. For a smooth disorder, with correlation
radius larger than the magnetic length, all diagrams are of the same
order of magnitude, but in this case magnetic phases, caused by
self--intersections of impurity lines, become small. The origin of
these phases lies in an uncertainty in the position of the center of
the Larmour orbit. Having the phases dropped, the entire
perturbation series can be summed up with the help of the Ward
identity, resulting in the Gaussian shape of the LL.\cite{P} 

For low LL numbers and short--range disorder, the magnetic phases in
diagrams are of the order of unity. A small parameter appears in the
problem  only  if the energy $\varepsilon$ (measured from the lowest LL)
is much larger than the LL width $\Gamma$, making possible a
calculation of the density of states (DOS) in the tails of LL. Such
calculations were carried out in the framework of the instanton
approach \cite{D,E,H,K,L,N,S} and  the tails were shown to be 
Gaussian. In the domain $\varepsilon \sim \Gamma$ the problem has no
small parameter and no simplifications are possible. However, for the
lowest LL, the exact DOS was found by Wegner \cite{F} for an arbitrary
ratio $\varepsilon/\Gamma$. Wegner has shown  that 
the diagrammatic expansion of the disorder--averaged Green
function, $G(\varepsilon)$, can be mapped onto that of the
zero--dimensional complex $\varphi ^4$--model with the partition
function $Z_0^{(1)}$ given by a simple integral

\begin{eqnarray}
\label{Z}
Z_0^{(1)}(\varepsilon , \Gamma) = \int d \varphi^{\ast} d \varphi 
\exp \Biggl[i\varepsilon \varphi^{\ast}\varphi
- \frac{\Gamma^2}{4}(\varphi^{\ast} \varphi)^2 \Biggr].
\end{eqnarray}          
The crucial observation made by Wegner was that the number of diagrams
for the disordered system, which are mapped onto a single graph of the
$\varphi ^4$--model, equals (up to an overall factor) the inverse
value of the diagram itself. The electron Green function is then given
by  

\begin{eqnarray}\label{gr}
G=-\frac{1}{2\pi l^2}\frac{\partial\ln Z_0^{(1)}}{\partial\varepsilon},
\end{eqnarray}
where $l$ is the magnetic length. Wegner has proved that coefficients
in front of $\Gamma^{n}$ in each side of this equation coincide.
Having a closed expression for $G(\varepsilon)$, Wegner obtained the
following formula for the DOS in the lowest LL

\begin{eqnarray}
\label{g}
g(\epsilon)=\frac{1}{2\pi^2l^2}\frac{2}{\sqrt \pi}
\frac{e^{\varepsilon^2/\Gamma^2}}
{1+\left(\frac{2}{\sqrt \pi}\int_0^{\varepsilon/\Gamma} dx
e^{x^2}\right)^2}. 
\end{eqnarray}
The magnetic field dependence of the  width $\Gamma$ is
$\Gamma \propto \sqrt B$. More precisely, for the correlator of the 
random potential $V({\bf r})$ of the form 
$\langle V({\bf r})V({\bf r}')\rangle =
\gamma\delta({\bf r}-{\bf r}')$,  one has 
$\Gamma =(\gamma/2\pi l^2)^{1/2}$.
An alternative derivation of Wegner's result was given by
Br\'{e}zin, Gross and Itzykson\cite{G} in the framework of 
functional--integral approach. 
  
Consider now a system consisting of two parallel two--dimensional
layers. In the absence of a disorder and magnetic field, a tunnel 
coupling between the layers would cause a splitting of size
quantization levels by an amount of $2t$, $t$ being the tunnel
integral. In a perpendicular magnetic field, the spectrum of the system
represents two staircases of LL shifted in energy by $2t$. Assume that
the field is strong, so that the cyclotron energy is much larger than
$t$. If a disorder is present in the layers, the resulting shape of
two adjacent LL's would depend on the  ratio $\Gamma/t$. If this ratio
is large, then the tunnel coupling does not play any role, so that the
DOS is twice the DOS in an individual layer, which is  given by
Eq.~(\ref{g}). In the opposite case, $t \gg \Gamma$, the peaks in the
DOS, corresponding to the symmetric and antisymmetric combinations of
size--quantization wave functions, are broadened independently. The
centers of the peaks are distanced by $2t$ and their shape is
described by Eq.~(\ref{g}) with the width\cite{W} $\Gamma/\sqrt{2}$. The
factor $1/\sqrt{2}$ appears because the effective random potential for
symmetric (antisymmetric) state is $[V_1({\bf r})\pm V_2({\bf r})]/2$,
where $V_1({\bf r})$ and $V_2({\bf r})$ are the random potentials in
the layers. If $\langle V_1({\bf r})V_2({\bf r}')\rangle =0$, the
correlator for each effective potential appears to be twice as small
as that for an individual layer.

In the  case $\Gamma \sim t$, calculation of the DOS in a
double--layer system seems to pose even harder problem than for a
single layer, since here the DOS represents a two--parametric
function, $g_{dl}(\varepsilon/\Gamma, t/\Gamma)$, with both arguments
of the order of unity. Nevertheless, as we demonstrate below, for the
lowest LL the exact DOS can be obtained in a closed form by
generalizing Wegner's approach.  Moreover, such a generalization can
be carried out for an arbitrary number of layers, and, in particular,
we consider the case when the number of layers is infinite
(superlattice). In the absence of a disorder, each LL in a
superlattice gives rise to a miniband of a width $4t$. Gradual
switching on a disorder first smears out the singularities in DOS 
at the edges of the miniband and then, as $\Gamma$ exceeds $t$,
transforms the DOS into a single peak, corresponding to an individual
layer. We derive an integral equation, the eigenvalue of which
determines the DOS in a superlattice, and trace this transformation by
solving it numerically.

Consider first the double--layer system. The Hamiltonian has the form

\begin{eqnarray}\label{ham}
\hat{H}=\sum_{i}\int d{\bf r}
\Biggl[\Biggl|\Biggl(\hat{\bf P}-\frac{e}{c}{\bf A}\Biggr)\psi_{i}\Biggr|^2
+\psi_i^{\ast}V_i\psi_i\Biggr] +
t\int d{\bf r}(\psi_1^{\ast}\psi_2+\psi_2^{\ast}\psi_1)
\end{eqnarray}
where ${\bf A}=(-By/2,Bx/2)$ is the vector-potential in the symmetric gauge
measured from the origin in both layers. It will be convenient to
include the last term in (\ref{ham}) into the definition of the free
Hamiltonian. Then, after projecting onto the lowest LL, the  free Green
function represents a $2\times 2$ matrix  

\begin{eqnarray}\label{green}
\hat{G}^{0}({\bf r,r'})=\frac{\hat{Q}}{2\pi l^2}
\exp\left[-\frac{({\bf r-r'})^2}{4l^2}+
\frac{i}{2l^2}({\bf r\times r'})\right]
\end{eqnarray}
with

\begin{eqnarray}\label{matr}
\hat{Q}=\left(\varepsilon -\hat{t}\right)^{-1},~~~ 
\hat{t}=
\left(\begin{array}{cc}
0 & t\\
t & 0
\end{array}\right).
\end{eqnarray} 
The perturbation expansion of the Green function, averaged over random
potentials $V_1$ and $V_2$, has the same diagrammatic representation
as for a single layer. Some of the first diagrams are shown in
Fig.~1(a) and (b). The solid lines correspond to $\hat{G}^{0}$ and 
the dashed lines
correspond to the correlator of the random potential. In contrast to
the single--layer case, solid lines carry indices, reflecting
the fact that electron can tunnel from one layer to another between
two successive scattering acts. Since the scattering retains an
electron in the same layer, the indices at the ends of each dashed line
are the same. Let us introduce projecting operators $\hat{\tau}_{i}$ as

\begin{eqnarray}
\label{proj}
\hat{\tau}_{1}=
\left(\begin{array}{cc}
1 & 0\\
0 & 0
\end{array}\right),~~~
\hat{\tau}_{2}=
\left(\begin{array}{cc}
0 & 0\\
0 & 1
\end{array}\right).
\end{eqnarray} 
Then the expressions corresponding to diagrams (a) and (b) can be
written as

\begin{eqnarray}\label{diag}
\hat{G}^{(1)}&=&F^{(1)}\Gamma^2\sum_i(\hat{Q}\hat{\tau}_{i}
\hat{Q}\hat{\tau}_{i}\hat{Q}),\nonumber\\
\hat{G}^{(2)}&=&F^{(2)}\Gamma^4\sum_{ij}(\hat{Q}\hat{\tau}_{i}
\hat{Q}\hat{\tau}_{j}\hat{Q}\hat{\tau}_{i}
\hat{Q}\hat{\tau}_{j}\hat{Q}),
\end{eqnarray} 
where $F^{(1)}$ and $F^{(2)}$ are spatial integrals. Similarly,
in any $n$th order diagram the spatial integrals are separated out as
factors in front of products of matrices $\hat{Q}$ and
$\hat{\tau}_{i}$, which are responsible for the energy dependence.
Important is that coefficients $F^{(n)}$ are {\em exactly the same as
those for a single-layer}. 

The mapping is carried out following Wegner's prescription: one
identifies pairs of points in a diagram connected by dashed
lines, and one gets a graph with four lines entering each vertex 
[see Fig.~1(c) and (d)]. In doing so, one obtains, in general, a set
of diagrams yielding the same 
graph. It is clear, however, that since assigning indices to the
lines does not alter in any way the topology of diagrams or graphs,
the number of diagrams in a set is the same for both single- and
double-layer cases. Moreover, one observes that the contractions of
matrices $\hat{\tau}_{i}$ precisely follow the identification of points
described above [as it can be seen, e.g., in Fig.~1(b) and (d)], so
that all the diagrams in such a set are equal. 
The fundamental relation, established by Wegner, is that for each
diagram in the set one has $F^{(n)}=1/{\cal N} s$, where ${\cal N}$ is
the number of diagrams in the set and $1/s$ is the symmetry factor of
the graph ($s$ is the number of permutations leaving graph invariant).
The latter factor is also unchanged by assigning indices to the graph. 
For example, the graph (d) is invariant under permutation of upper
and lower lines so its symmetry factor is 1/2 in both cases. Thus, the
contribution of the set, being proportional to ${\cal N}F^{(n)}$, is 
${\cal N}$--independent, and the problem again reduces to the
zero-dimensional field theory. The remaining question is whether
matrix products of type (\ref{diag}) can be generated in the
perturbation expansion of some generalized $\varphi^4$--model.
Our main observation is that the model with the partition function 

\begin{eqnarray}
\label{Z2}
Z_0^{(2)} = \int d\Phi^{\ast}d\Phi 
\exp \Biggl[i\Phi^{\ast}\hat{Q}^{-1}\Phi
-\frac{\Gamma^2}{4}\sum_{i}(\Phi^{\ast} \tau_i\Phi)^2 \Biggr],
\end{eqnarray}
accomplishes this task. Here $\hat{Q}$ and $\hat{\tau}_{i}$ are
matrices defined by (\ref{matr}), and $\Phi$ is a {\em two-component}
complex field: $\Phi=(\varphi_1,\varphi_2)$. Indeed, the $n$th order
term in the expansion of exponent (\ref{Z2}) in terms of $\Gamma^2$
represents a product of $2n$ matrices $\hat{\tau}_{i}$ (with all
pairwise contractions) separated by $2n$ products of the form
$\Phi\Phi^{\ast}$. Then the gaussian integral over $\Phi$ inserts the
``Green function'' $\hat{Q}= -i\langle \Phi\Phi^{\ast}\rangle$ in
place of each pair of fields $\Phi$ and $\Phi^{\ast}$, with all
possible contractions between them yielding all the $n$th order graphs
with appropriate symmetry factors.

Having the mapping established, the DOS in the double-layer system can
be calculated directly from (\ref{Z2}). It is also instructive 
to rewrite $Z_0^{(2)}$ in a different form. First, we decouple the
quartic term in the exponent of (\ref{Z2}),

\begin{eqnarray}\label{action}
iS=\sum_{j=1,2}\Biggl[i\varepsilon\varphi_j^{\ast}\varphi_j
-\frac{\Gamma^2}{4}(\varphi_j^{\ast} \varphi_j)^2\Biggr]
-it(\varphi_1^{\ast}\varphi_{2}+\varphi_{2}^{\ast}\varphi_1),
\end{eqnarray} 
with the help of gaussian integral over a pair of auxiliary variables. 
Performing the remaining integral over $\varphi_i$ we then obtain

\begin{eqnarray}\label{ZZ2}
Z_0^{(2)}=\frac{(i\pi)^2}{\pi\Gamma^2}
\int_{-\infty}^{\infty}\int_{-\infty}^{\infty}
\frac{d\lambda_1 d\lambda_2}
{(\varepsilon+\lambda_1)(\varepsilon+\lambda_2)-t^2}
\exp\left(-\frac{\lambda_1^2}{\Gamma^2}
-\frac{\lambda_2^2}{\Gamma^2}\right).
\end{eqnarray}
From the form (\ref{ZZ2}), the both limiting cases of large and small
$t$ are evident. For small $t$, the partition function factorizes,
$Z_0^{(2)}=\left(Z_0^{(1)}\right)^2$, yielding 
twice the DOS (\ref{g}). For $t \gg \Gamma$ the characteristic values
of $\lambda_1$, $\lambda_2$ in (\ref{ZZ2}), being of the order of
$\Gamma$, are much smaller than $t$. This allows to neglect the
product $\lambda_1 \lambda_2$ in the denominator; $Z_0^{(2)}$ is not
small only if $(\varepsilon - t)\sim \Gamma$ or 
$(\varepsilon + t)\sim \Gamma$. In both cases one should introduce new
variables $\mu_1=\lambda_1+\lambda_2$ and $\mu_2=\lambda_1-\lambda_2$.
Then the integration over $\mu_2$ would contribute a factor
$\sqrt{2\pi}\Gamma$, and the integral over $\mu_1$ would reproduce
Wegner's result with the width $\Gamma/\sqrt{2}$, as discussed above.  
The evolution of DOS between two limits, calculated from (\ref{gr})
with $Z_0^{(2)}$, is shown in Fig.~2.

Let us now turn to a superlattice. The partition function (\ref{Z2})
can be straightforwardly generalized to a multilayer system, and for
$n$ layers with nearest-neighbor tunneling it takes the form

\begin{eqnarray}
\label{Z_n}
Z_0^{(n)}(\varepsilon , \Gamma) = \int \prod_{i=1}^{n}
d \varphi_i^{\ast} d \varphi_i
\exp \Biggl[i\varepsilon \sum_{j=1}^{n}\varphi_j^{\ast}\varphi_j -
it\sum_{j=1}^{n-1}(\varphi_j^{\ast}\varphi_{j+1}
+\varphi_{j+1}^{\ast}\varphi_j)
-\frac{\Gamma^2}{4}\sum_{j=1}^{n}(\varphi_j^{\ast} \varphi_j)^2 \Biggr].
\end{eqnarray}
We are interested in the asymptotic behavior of $Z_0^{(n)}
(\varepsilon , \Gamma)$ as $n\rightarrow\infty$. For this purpose we
employ a method  similar to the transfer--matrix method in
the theory of 1D spin chains. Note that the expression (\ref{Z_n})
for $Z_0^{(n)}$ can be rewritten as 

\begin{eqnarray}
\label{I}
Z_0^{(n)}(\varepsilon , \Gamma) = \int d \varphi^{\ast} d \varphi
\exp \Biggl[i\varepsilon \varphi^{\ast}\varphi
-\frac{\Gamma^2}{4}(\varphi^{\ast} \varphi)^2 \Biggr]
I_n(\varphi^{\ast},\varphi),
\end{eqnarray} 
where $I_1=1$ and  the functions $I_n(\varphi^{\ast},\varphi)$ satisfy
the following recurrence relation

\begin{eqnarray}
\label{T}
I_{n+1}(\varphi^{\ast},\varphi)=\hat{T}_{\varepsilon,\Gamma}I_{n}= 
\int d \varphi_1^{\ast} d \varphi_1
\exp \Biggl[i\varepsilon \varphi_1^{\ast}\varphi_1
-\frac{\Gamma^2}{4}(\varphi_1^{\ast} \varphi_1)^2 
-it(\varphi^{\ast}\varphi_1+\varphi_1^{\ast}\varphi)\Biggr]
I_{n}(\varphi_1^{\ast},\varphi_1).
\end{eqnarray}
Consider now the eigenvalues, $\lambda^{(k)}(\varepsilon,\Gamma)$, and
eigenfunctions, 
$\Omega^{(k)}_{\varepsilon,\Gamma}(\varphi^{\ast},\varphi)$, 
of the operator $\hat{T}_{\varepsilon,\Gamma}$: 
$\hat{T}_{\varepsilon,\Gamma}\Omega^{(k)}=\lambda^{(k)}\Omega^{(k)}$. 
Assume that $\lambda^{(0)}$ has the maximal absolute value. 
Then in the limit $n\rightarrow\infty$, $Z_0^{(n)}$ will behave as
$(\lambda^{(0)})^n$. Hence, the DOS per layer  in a superlattice can
be expressed through $\lambda^{(0)}(\varepsilon,\Gamma)$ in the 
following way  

\begin{eqnarray}
\label{sl}
g_{sl}(\varepsilon,\Gamma)=\frac{1}{2\pi^2 l^2}
\mbox{Im} \lim_{n\rightarrow\infty}
\frac{1}{n}\frac{\partial\ln Z_0^{(n)}}{\partial\varepsilon}
=\frac{1}{2\pi^2 l^2}
\mbox{Im}\frac{\partial\ln \lambda^{(0)}(\varepsilon,\Gamma)}
{\partial\varepsilon}.
\end{eqnarray}
Thus, we have reduced the calculation of DOS to the solution of an
integral equation  

\begin{eqnarray}
\label{equ}
\lambda^{(0)}(\varepsilon,\Gamma)
\Omega_{\varepsilon,\Gamma}^{(0)}(\varphi^{\ast},\varphi)= 
\int d \varphi_1^{\ast} d \varphi_1
\exp \Biggl[i\varepsilon \varphi_1^{\ast}\varphi_1
-\frac{\Gamma^2}{4}(\varphi_1^{\ast} \varphi_1)^2 
-it(\varphi^{\ast}\varphi_1+\varphi_1^{\ast}\varphi)\Biggr]
\Omega_{\varepsilon,\Gamma}^{(0)}(\varphi_1,\varphi_1^{\ast}).
\end{eqnarray}
Consider first the case of a weak disorder, $\Gamma\rightarrow 0$. One
can check that eigenfunctions of $\hat{T}_{\varepsilon,0}$ in this
case have the form

\begin{eqnarray}
\label{eigen}
\Omega_{\varepsilon,0}^{(p,m)}=e^{im\alpha-
(i\varepsilon+\sqrt{4t^2-\varepsilon^2})R^2/2}
\left(R^2\sqrt{4t^2-\varepsilon^2}\right)^{m/2}
L_p^{m}\left(R^2\sqrt{4t^2-\varepsilon^2}\right),
\end{eqnarray}
where $R$ and $\alpha$ are, respectively, the absolute value and the
phase of  $\varphi$, and $L_p^m(x)$ is the Laguerre polynomial.
The corresponding eigenvalues, $\lambda^{(p,m)}(\varepsilon,0)$, are 
equal to

\begin{eqnarray}
\label{0}
\lambda^{(p,m)}(\varepsilon,0)=\frac{\pi i}{t}
\left(\frac{2it}{i\varepsilon-\sqrt{4t^2-\varepsilon^2}}\right)^{2p+m+1},
\end{eqnarray}
where for $|\varepsilon|>2t$ the square root is defined as
$i^{-1}\mbox{sgn}(\varepsilon)\sqrt{\varepsilon^2-4t^2}$.
Outside the interval $|\varepsilon|<2t$, the phases  of eigenvalues
(\ref{0}) have no energy dependence, supporting  the obvious
observation that the DOS is zero outside the miniband. Within the
miniband, all $\lambda^{(p,m)}(\varepsilon,0)$ have the same  absolute
value. This is a manifestation of the fact that for a large but finite
number of layers the DOS in the absence of disorder represents a set
of delta--peaks. However, with arbitrary weak disorder present, only
the eigenvalue  $\lambda^{(0,0)}(\varepsilon,0)$ will survive in the
limit $n\rightarrow\infty$,  yielding the familiar result 

\begin{eqnarray}
\label{g0}
g_{sl}(\varepsilon,0)= \frac{1}{2\pi^2 l^2}
\mbox{Im}\frac{\partial\ln \lambda^{(0,0)}(\varepsilon,0)}
{\partial\varepsilon}=\frac{1}{2\pi^2 l^2 \sqrt{4t^2-\varepsilon^2}}.
\end{eqnarray}
Assume now that the disorder is finite but $\Gamma\ll t$. It is
convenient to formally rewrite  Eq.~(\ref{equ}) in the following form

\begin{eqnarray}
\label{form}
\lambda^{(0)}(\varepsilon,\Gamma)\Omega^{(0)}_{\varepsilon,\Gamma} 
=\hat{T}_{\varepsilon,\Gamma}\Omega^{(0)}_{\varepsilon,\Gamma}
=\frac{1}{\sqrt{\pi}\Gamma}\int_{-\infty}^{\infty}dE 
\exp\left[-\frac{(E-\varepsilon)^2}{\Gamma^2}\right]\hat{T}_{E,0}
\Omega^{(0)}_{\varepsilon,\Gamma}.
\end{eqnarray}  
For small $\Gamma$, only the energies $E$ close to $\varepsilon$
contribute to the integral (\ref{form}). This suggests to  start the
iteration procedure by substituting, as a zero approximation, the  
$m=p=0$ eigenfunction of $\hat{T}_{E,0}$,
$\Omega_{E,0}^{(0,0)}= 
\exp\left[-\left(iE +\sqrt{4t^2-E^2}\right)R^2/2\right]$,
into the right--hand side. This generates the first approximation for
the function $\Omega^{(0)}_{\varepsilon, \Gamma}$

\begin{eqnarray}
\label{tilde}
\tilde \Omega^{(0)}_{\varepsilon,\Gamma}=
\frac{1}{\sqrt{\pi}\Gamma\lambda^{(0)}(\varepsilon,\Gamma)}
\int_{-\infty}^{\infty}dE
\exp\left[-\frac{(E-\varepsilon)^2}{\Gamma^2}\right]
\lambda^{(0,0)}(E,0)\Omega_{E,0}^{(0,0)}.
\end{eqnarray}
Substituting this function back into $(\ref{form})$, we obtain
\begin{eqnarray}
\label{ansatz}
\hat{T}_{\varepsilon,\Gamma}\tilde{\Omega}^{(0)}_{\varepsilon,\Gamma}=
\frac{1}{\pi\Gamma^2\lambda^{(0)}(\varepsilon,\Gamma)}
\int_{-\infty}^{\infty} dE \int_{-\infty}^{\infty}
dE'\exp\left[-\frac{(E-\varepsilon)^2}{\Gamma^2}-
\frac{(E'-\varepsilon)^2}{\Gamma^2}\right]
\lambda^{(0,0)}(E',0)\hat{T}_{E,0}
\Omega_{E',0}^{(0,0)}.
\end{eqnarray}
Note now, that  $\Omega_{E',0}^{(0,0)}$ as a function of $E'$ changes
significantly on the scale $E'\sim t$. On the other hand, exponential 
factors in (\ref{ansatz}) enforce the difference between $E$ and $E'$ be
of the order of $\Gamma$. This allows to replace $\Omega_{E',0}^{(0,0)}$
by $\Omega_{E,0}^{(0,0)}$ under the integral. Then we immediately 
observe that the right--hand side takes the form
$\tilde{\lambda}^{(0)}(\varepsilon,\Gamma)
\tilde{\Omega}^{(0)}_{\varepsilon,\Gamma}$ with  

\begin{eqnarray}
\label{last}
\tilde{\lambda}^{(0)}(\varepsilon,\Gamma)=
\frac{2\pi}{\sqrt{\pi}\Gamma}\int_{-\infty}^{\infty} \frac{dE}
{\sqrt{4t^2-E^2}-iE}\exp\left[-\frac{(E-\varepsilon)^2}{\Gamma^2}\right].
\end{eqnarray}
In other words, for small $\Gamma$ the function
$\tilde{\Omega}^{(0)}_{\varepsilon,\Gamma}$
satisfies Eq.~(\ref{form}), yielding the eigenvalue (\ref{last}). 

In principle, to assess the region of large $\Gamma$ one should keep
iterating Eq.~(\ref{form}). However, as  we have established
numerically,  the function $\tilde{\Omega}^{(0)}$ is already a very good
approximation for $\Omega^{(0)}$ and $\tilde{\lambda}^{(0)}$ is a very good
approximation for $\lambda^{(0)}$ not only for small, but for arbitrary
ratio $\Gamma/t$. The reason for this is the following.  As $\Gamma/t$  
increases, one should reproduce Wegner's result, which corresponds to
$t=0$ and, hence, $\Omega^{(0)}=const$ in Eq.~(\ref{equ}). On the  other
hand, it is easy to see that $\tilde{\Omega}^{(0)}$ turns to constant as
$t\rightarrow 0$, and that in this limit $\tilde{\lambda}^{(0)}$ turns to 
$Z_0^{(1)}$. Thus, Eq.~(\ref{last}) is exact in both limits. The
numerical results for the DOS, obtained by substituting
$\tilde{\lambda}^{(0)}$ into (\ref{sl}), are shown in Fig.~3. We see that the
miniband is completely destroyed as $\Gamma/t$ exceeds $1.4$.

Note in conclusion, that  a decade ago there was a significant interest
in the study of transport phenomena in multilayer systems in a strong 
magnetic field (see, e.g., Refs.\onlinecite{T} and\onlinecite{U}).
In the recent publications\cite{V,W} this interest was renewed.
The question  of interest is how a transition from a purely 2D to
three-dimensional behavior of  the conductivity occurs with
increasing $t$. As was shown in Ref.\onlinecite{X}, the structure 
of electronic states  in a multilayer system can be efficiently tuned
by tilting magnetic field. In this case the role of the parallel
component of $B$ reduces to the suppression of the interlayer tunneling.

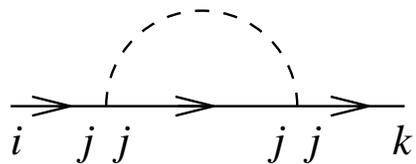
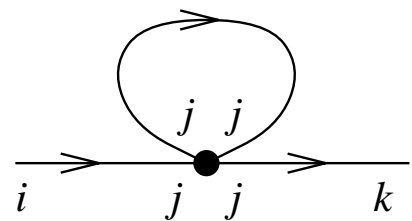
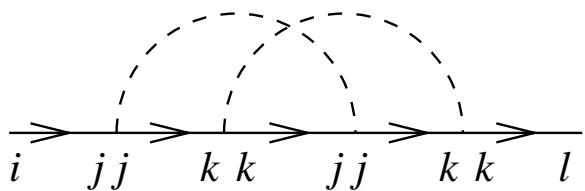
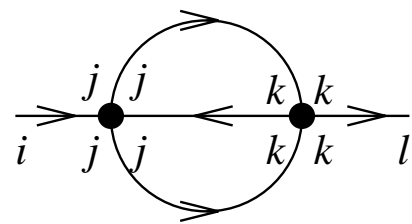
\begin{figure}
\caption{First (a) and second (b) order diagrams for the Green function
mapped on graphs (c) and (d), respectively.}
\end{figure}

\begin{figure}
\caption{The DOS per layer for a double-layer system in units of
$g_1=(2\pi\l^2)^{-1}\Gamma^{-1}$ for values of
$t/\Gamma$=0.0 (highest curve), 0.4, 0.6, 0.8, 1.0, 1.2, 1.4, 1.6, 1.8,
2.0, 2.2, 2.4, 2.6, 2.8, and 3.0.}
\end{figure}

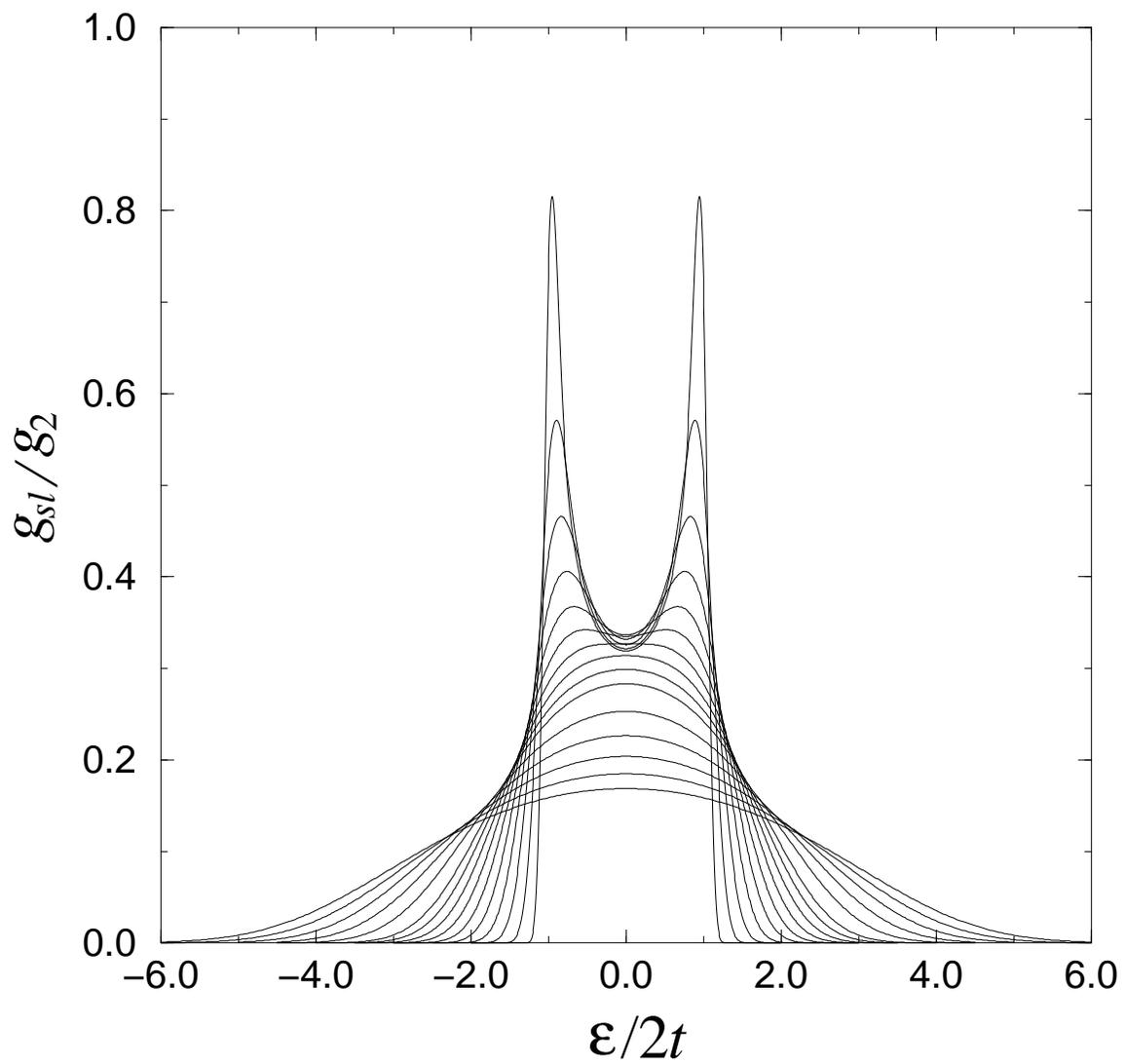
\begin{figure}
\caption{The DOS per layer for a superlattice  in units of 
$g_2=(2\pi\l^2)^{-1}(2t)^{-1}$ for values of
$\Gamma/2t$=0.1 (highest curve), 0.2, 0.3, 0.4, 0.5, 0.6, 0.7, 0.8, 0.9,
1.0, 1.2, 1.4, 1.6, 1.8, and 2.0.}
\end{figure}
\end{document}